\documentclass[reqno,12pt]{article}
\usepackage{amsmath,amssymb,amsfonts,epsfig,graphicx,euscript}
\usepackage{graphics}
\usepackage{graphicx}
\usepackage{dcolumn}
\usepackage{bm}
\usepackage{amssymb}
\usepackage{amsmath,bm}
\usepackage{amsfonts}
\usepackage{slashed}
\newcommand{\bse}{\begin{subequations}}
\newcommand{\ese}{\end{subequations}}
\newcommand{\be}{\begin{equation}}
\newcommand{\ee}{\end{equation}}
\newcommand{\bea}{\begin{eqnarray}}
\newcommand{\eea}{\end{eqnarray}}
\newcommand{\ba}{\begin{array}}
\newcommand{\ea}{\end{array}}

\begin{document}

\begin{flushright}
\end{flushright}
\begin{flushright}
\end{flushright}
\hfill%
\begin{center}

{\LARGE {\sc Tensor Fields on Self-Dual Warped $AdS_3$
Background}}

\bigskip
{ Mohammad A. Ganjali\footnote{ganjali@theory.ipm.ac.ir}, Sadegh
Sadeghian
} \\
{Department of Physics, Kharazmi University,\\P. O. Box
31979-37551, Tehran, Iran}
\\

\end{center}

\bigskip
\begin{center}
{\bf { Abstract}}\\
\end{center}
Considering rank s fields obey first order equation of motion, we
study the dynamics of such fields in a 3 dimensional self-dual
space-like warped $AdS_3$ black hole background. We argue that in
this background, symmetric conditions and gauge constraint can
not be satisfied simultaneously. Using new suitable constraint,
we find the exact solutions of equation of motion. Then, we obtain
the quasi-normal modes by imposing appropriate boundary condition
at horizon and infinity.

\newpage
\section{Introduction}
Quantum Gravity as a theory of consistently quantized symmetric
massless rank 2 field is one of the most challenging subjects for
high energy physicist  during the last decade. An outstanding
efforts have been done for overcoming this problem and some
elegant and interesting theories such as super-symmetric field
theory, string theory, etc, invented and many insights gained but
the problem is unsolved completely up to now.

The recently attracted subject in this field was studying the
gravity in three dimensions. Although, it seems that Einstein
gravity is trivial in three dimensions\cite{Witten:2007kt} but,
by adding higher correction to usual Einstein action with
cosmological constant, one obtains theories which have
propagating degrees of freedom\cite{Li:2008dq}.

One of such theories is Topologically Massive Theory(TMG) which
its higher derivative terms are gravitational Chern-Simons
term\cite{Deser:1982vy,Deser:1981wh}
 \bea
I_{TMG}=\frac{1}{16\pi G}\left(I_{EH}+\frac{1}{\mu}I_{CS} \right)
 \eea
 with $\mu$ is coupling constant and $I_{EH}$ includes the cosmological term
 $-2\Lambda=\frac{2}{l^2}$.

TMG admits maximally symmetric solutions $AdS_3$ with
$SL(2,\mathbb{R})\times SL(2,\mathbb{R})$
symmetry\cite{Anninos:2008fx} and BTZ black
hole\cite{Banados:1992wn} and also solutions with fewer
symmetries known as warped $AdS_3$ and its corresponding black
holes\cite{Vuorio:1985ta}. Due to hidden conformal symmetries of
the propagating modes on the warped background, it was
conjectured  that such gravitational warped solution has a dual
conformal field theory with the corresponding central
charges\cite{Anninos:2008fx}. Motivated by this observations,
many works has been done in this line for better understanding
the duality and other related topics of quantum gravity and black
holes in three dimensions\cite{Orlando:2010ay}.

Self-dual warped $AdS_3$ solution is also solution of
TMG\cite{Chen:2010qm}. It may be defined as real line fibrations
over $AdS_2$ which preserves a single $SL(2,\mathbb{R})$ isometry
and a non-compact $U(1)$ isometry generated by translation along
the fibre coordinate.

Such background has some very interesting properties and its
several aspects such as geometrical properties, thermodynamics,
its CFT dual and etc, has been studied. Especially, using an
algebraic way, the author of \cite{Chen:2010ik} were found the
quasi-normal modes of scalar, vector and tensor perturbations for
$AdS_3, BTZ$ and warped $AdS_3$ background. This calculation has
been done using other methodes such as finding the exact
solution\cite{Birmingham:2001pj}. For example, in
\cite{Birmingham:2001hc,Datta:2011za}, the authors were able to
find the exact solution in BTZ black hole background for any
general integer or half-integer rank s field and found the
quasi-normal modes.

In fact, till recently, because of some no-go theorems
\cite{Weinberg:1980kq}, efforts typically was focused on the
dynamics of fields with spin lower than 2 in gravitational
backgrounds.

But, due to the work of \cite{Vasiliev:1989mw} the Higher Spin
fields were entered to the game and seems that have somehow
serious rolls in fundamental physics. Various aspects of theories
with higher spin fields such as finding a Lagrangian formalism,
studying the conserved current and so on have been studied, see
for example\cite{Tyutin:1997yn}.

Our aim in this paper is to continue this research line by
studying with details the constraint in which a rank s field
should satisfy in this background and obtaining the exact
solutions of tensor mode perturbations and finally, finding the
quasi-normal modes. We will also study modes with spin greater
than 2(higher spin fields) on this background.

For self-dual warped $AdS_3$, we suppose that the main equation in
which higher spin fields should obey is
\begin{equation}\label{leom}
\epsilon _{\mu}^{\ \alpha\beta} \nabla _\alpha \Phi
_{\beta\nu_2\nu_3\cdots\nu_{s}}=-m \Phi
_{\mu\nu_2\nu_3\cdots\nu_{s}}.
\end{equation}
Due to lower symmetry in this background, so is not locally $AdS$
and it was shown in \cite{Anninos:2009zi} that such a simple
first order equation can not be carried out for a metric
perturbation but, one can still consider the above equation for a
massive spin s field in the given background\cite{Chen:2010ik}.

Nevertheless, in spite of BTZ background\cite{Datta:2011za}, such
a linear equation does not imply the following simple form of
second order Klein-Gordon equation and gauge condition for all
components of a spin s field\footnote{Note that the symmetric
rank s field obey traceless condition as $ g^{\mu\nu}
\Phi_{\mu\nu \mu_3 \cdots \mu_s}=0$}
\cite{Metsaev:2003cu,Buchbinder:2006ge}
\begin{eqnarray}\label{eom}
( \nabla^2  -m^2_s ) \Phi_{\mu_1\mu_2\mu_2 \cdots \mu_s} &=&0,  \\
 \nabla^\mu \Phi_{\mu \mu_2 \cdots \mu_s} &=& 0.
\end{eqnarray}
 In fact the situation is worse noting that the linear equation and the gauge constraint
 for a symmetric field satisfied only for zero mode(s-wave).

However, we will show that by imposing a suitable weaker
condition, one can write the equation for some components in the
Klein-Gordon form and can solve them exactly. In particular, for
$\Phi_{\theta\theta\cdots\theta}$ and $\Phi_{t\theta\cdots\theta}$
components, the above equations can be modified slightly and solve
it exactly. Then using the linear equation(\ref{leom}) and some
contiguous relations between the hyper-geometric functions, one
can obtain the solution for the other components and quasi normal
modes and also the left and right conformal weight.

The paper organized as follows. In section 2, we briefly
introduce the three dimensional self dual Warped $AdS_3$
background. In the next section, we discuss about the equation of
motion and gauge constraint of higher spin fields in this
background and will see that all components do not satisfy
Klein-Gordon equation and harmonic constraint. In section four,
we find the exact solutions for spin 2 case for two components
$h_{\theta\theta}$ and $h_{t\theta}$ and then using the first
order equation (\ref{leom}) we find the solution for $h_{\theta
t}$ and $h_{tt}$. After that, by imposing Dirichlet Boundary
condition, we find the quasi-normal modes for massive spin 2
field. In section 5 we generalize our computations for higher
spin field and find the quasi normal modes.
\section{\textsl{Self Dual Warped $AdS_3$} Background}
In this section, we briefly introduce self-dual warped AdS3
solution. \textsl{Self dual warped $AdS_3$} is a solution of
equation of motion of Topological Massive Gravity(TMG) in three
dimensions\cite{Chen:2010qm}.
 The metric is given by
 \begin{eqnarray}\label{selfmetric}
 ds^2&=&\frac{l^2}{\nu^2+3}\left(-\left(x-x_+\right)\left(x-x_-\right)d\tau^2
 +\frac{1}{\left(x-x_+\right)\left(x-x_-\right)}dx^2\right.\nonumber\\
 &&+\left.\frac{4\nu^2}{\nu^2+3}\left(\alpha d\theta+
 \frac{1}{2}\left(2x-x_+-x_-\right)d\tau\right)^2
 \right)\;,
 \end{eqnarray}
 where $x_+$ and $x_-$ are the
 outer and inner horizons radius respectively.

In fact, the warped vacua of TMG classified into three spacelike,
timelike and null warped types. Classification depends on the
whether the norm of the Killing vector generating the U(1)
isometry is positive, negative or zero. First two types can also
be classified as stretched ($\mu l > 3$) or squashed ($\mu l <
3$) depending on the magnitude of the warp factor. These
background spacetimes are not locally $AdS_3$.

Quotients of warped AdS3 along various Killing directions may
give rise to black holes\cite{Anninos:2008fx}. Black hole
solutions free of closed timelike curves (CTCs) can only be found
in spacelike stretched and null warped $AdS3$. Self-dual solutions
in $AdS_3$ is quotients of spacelike warped $AdS_3$ along the
U(1). Such geometries have Killing horizons and no
CTCs\cite{Coussaert:1994tu} .

It is shown in \cite{Chen:2010qm} that under the consistent
boundary condition, the U(1) isometry is enhanced to a Virasoro
algebra with nonvanishing left central charge while the
$SL(2,\mathbb{R})$ isometry becomes trivial with the vanishing
right central charge,
 \bea
c_L=\frac{4\nu l}{\nu^2+3},\;\;\;\;\;\;\;\;c_R=o.
 \eea
It is conjectured that the self-dual warped AdS3 black hole is
dual to a two dimensional chiral CFT, which provides an example of
warped AdS/CFT dual.

The left and right temperatures of CFT are defined
by\cite{Chen:2010qm,Chen:2009hg}
 \bea
T_L=\frac{\alpha}{2\pi
l},\;\;\;\;\;\;\;\;\;T_R=\frac{x_+-x_-}{4\pi l}
 \eea
The mass $M$ and angular momentum $J$ of this black hole
 are given by
 \begin{eqnarray}
 M=0\;,\;\;\;J=\frac{(\alpha^2-1)\nu}{6G(\nu^2+3)}\;.
 \end{eqnarray}

The angular velocity of the event horizon $\Omega_H$ and the
Bekenstein-Hawking entropy $S_{BH}$ of this solution are
respectively given by
 \begin{eqnarray}\label{temprature}
 \Omega_H&=&-\frac{x_+-x_-}{2\alpha}\;,\nonumber\\
 S_{BH}&=&\frac{2\pi\alpha\nu}{3G(\nu^2+3)}\;.
 \end{eqnarray}

Note also that, this solution is asymptotic to the spacelike
warped $AdS_3$ spacetime and under a suitable coordinate
transformation
the metric of self-dual warped $AdS_3$ black hole can be
transformed to the metric of spacelike warped AdS$_3$ spacetime
 \begin{eqnarray}
 ds^2=\frac{1}{\nu^2+3}\left(-\cosh^2\sigma dv^2+d\sigma^2
 +\frac{4\nu^2}{\nu^2+3}(du+\sinh\sigma dv)^2\right)\;.
 \end{eqnarray}
At the end, let us mention that since the self-dual warped
$AdS_3$ is not locally $AdS$ its curvature, Riemann and Ricci
tensor can not be written in the usual form in terms of
metric\footnote{In a maximally symmetric space one has
$R_{\mu\nu\lambda\theta}=g_{\mu\theta}g_{\nu\lambda}-g_{\mu\lambda}g_{\nu\theta}=\epsilon_{\mu\nu\alpha}\epsilon_{\lambda\theta\beta}
G^{\alpha\beta}$ where $G^{\alpha\beta}$ is Einstein tensor and
we have defined  $\epsilon^{xt\theta}=+\frac{1}{\sqrt{-g}}$.}. In
fact, one can see that for background (\ref{selfmetric})
 \bea
R_{\mu\nu\lambda\theta}=R^0_{\mu\nu\lambda\theta}+r_{\mu\nu\lambda\theta},
 \eea
where
 \bea
R^0_{\mu\nu\lambda\theta}&=&\frac{\lambda^2}{4}(g_{\mu\theta}g_{\nu\lambda}-g_{\mu\lambda}g_{\nu\theta}),\cr
r_{rtrt}&=&1-\lambda^2,\;\;\;\;\;\;r_{other}=0
 \eea
and we have defined $\lambda^2=\frac{4\nu^2}{\nu^2+3}$. For later
use, we also define $x_0=\frac{x_++x_-}{2}$ and
$\bar{x}_0=\frac{x_+-x_-}{2}$.

Moreover, there are some relations between metric components and
Christoffel symbols as\footnote{These relations can be recasted
as $\mathop g\nolimits^{xx} \mathop \Gamma \nolimits_{x\nu }^\mu
\mathop { = - \mathop g\nolimits^{\mu\beta} \Gamma
}\nolimits_{\beta\nu}^x$.}
\begin{equation}\label{gammarelation}
\begin{array}{l}
\mathop g\nolimits^{xx} \mathop \Gamma \nolimits_{x\theta }^\theta
\mathop { =  - \mathop g\nolimits^{\theta t} \Gamma
}\nolimits_{\theta t}^x,\;\;\;\;\;\; \mathop g\nolimits^{tt}
\mathop \Gamma \nolimits_{\theta t}^x  =
 - \mathop g\nolimits^{xx} \mathop \Gamma \nolimits_{x\theta }^t, \\
\mathop g\nolimits^{xx} \mathop \Gamma \nolimits_{xt}^\theta   = -
\mathop g\nolimits^{\theta \theta } \mathop \Gamma
\nolimits_{\theta t}^x
- \mathop g\nolimits^{\theta t} \mathop \Gamma \nolimits_{tt}^x \\
\mathop g\nolimits^{xx} \mathop \Gamma \nolimits_{xt}^t  =  -
\mathop g\nolimits^{\theta t} \mathop \Gamma \nolimits_{\theta
t}^x  - \mathop g\nolimits^{tt} \mathop \Gamma \nolimits_{tt}^x.
\end{array}
\end{equation}
These relations are very useful and important in finding the final
equations. In fact, our discussion about the equations and quasi
normal modes can be generalized to any other metrics in 3
dimensions in which $g_{x\mu}=0,\mu\neq x$ and satisfied
(\ref{gammarelation}).

Without loose of generality, we will set $\frac{l^2}{\nu^2+3}=1$
and $\alpha=1$.
\section{Fields on Self-dual Warped $AdS_3$}
Massive integer spin $s$ fields in $AdS_3$ spaces are realized by
totally symmetric tensors of rank $s$ satisfying the following
equation of motion, gauge condition and traceless condition
as\cite{Metsaev:2003cu,Buchbinder:2006ge}
\begin{eqnarray}
\label{eom-s} ( \nabla^2  -m^2_s ) \Phi_{\mu_1\mu_2\mu_2 \cdots \mu_s} &=&0,  \\
 \nabla^\mu \Phi_{\mu \mu_2 \cdots \mu_s} &=& 0, \\
 g^{\mu\nu} \Phi_{\mu\nu \mu_3 \cdots \mu_s} &=& 0.
\end{eqnarray}
Here,  $\Phi_{\mu_1 \mu_2 \cdots \mu_s}$ is a totally symmetric
rank $s$ tensor and for $AdS_3$ the mass of the field is given by
$m_s^2 = s(s-3)+M^2$. The first term is the mass that exists due
to the curvature of $AdS_3$.

The above set of equations in a maximally symmetric space  are
equivalent with the following first order equation
\cite{Datta:2011za}
\begin{equation}\label{lleom}
\epsilon _{\mu}^{\ \alpha\beta} \nabla _\alpha \Phi
_{\beta\nu_2\nu_3\cdots\nu_{s}}=-m \Phi
_{\mu\nu_2\nu_3\cdots\nu_{s}}.
\end{equation}
where $m^2=M^2+(s-1)^2$.

However, it was shown\cite{Anninos:2009zi} in self-dual warped
$AdS_3$ space, where the background is not locally $AdS$, one can
not rewrite the field equations for metric perturbations in a
simple form as (\ref{leom}). But, one can yet consider any
massive tensor field $\Phi_{\nu_1\nu_2\nu_3\cdots\nu_{s}}$ obeys
the above first order equation\cite{Chen:2010ik}. Adopting
(\ref{leom}) for a tensor field
$\Phi_{\nu_1\nu_2\nu_3\cdots\nu_{s}}$, one may search a set of
equations similar to (\ref{eom}). Unfortunately, because of the
reasons which will be presented bellow, the gauge constraint and
the tracelessness condition do not satisfied for a general
\textsl{"symmetric"} rank s tensor field
$\Phi_{\nu_1\nu_2\nu_3\cdots\nu_{s}}$. Indeed, the Klein-Gordon
equation can not be written in the simple form (\ref{eom}) for
all components of a given field.

The argument is as follows. Let us for simplicity consider a rank
2 field $h_{\mu\nu}$ and suppose that is totally symmetric and
obeys the equation (\ref{leom}). It means that that
 \bea\label{tracelessness}
0&=&\epsilon^{\mu\nu\rho}\epsilon_{\mu}^{\;\;\alpha\beta}\nabla_{\alpha}h_{\beta\nu}\cr
&=&-g^{\rho\beta}\nabla^{\alpha}h_{\beta\alpha}+g^{\rho\alpha}\nabla_{\alpha}g^{\nu\beta}h_{\beta\nu}
 \eea
In the other hand, using (\ref{leom}) for $h_{rr}$ one obtains
 \bea\label{rr}
-mh_{rr}&=&g_{rr}\epsilon^{rt\theta}\left(\nabla_{t}h_{\theta
r}-\nabla_{\theta}h_{tr}\right)\cr
&=&g_{rr}\epsilon^{rt\theta}\left(\nabla_{t}h_{r\theta}
-\nabla_{\theta}h_{rt}\right)\cr &=&
g_{rr}\epsilon^{rt\theta}\left(\nabla_{r}(h_{t\theta}-h_{\theta t
})-mg_{rr}\epsilon^{rt\theta}(g_{tt}h_{\theta\theta}-g_{t\theta}h_{t\theta}-g_{\theta
t}h_{\theta t}+g_{\theta\theta}h_{tt})\right)\cr
&=&\frac{m}{g^{rr}}\left(0+g^{\theta\theta}h_{\theta\theta}+2g^{t\theta}h_{t\theta}+g^{tt}h_{tt}\right)
 \eea
In the third line of (\ref{rr}) we have used (\ref{leom}) for
some other components\footnote{Notice that (\ref{leom}) can be
written as
$$\nabla_{\lambda}h_{\alpha\nu}-\nabla_{\alpha}h_{\lambda\nu}=m\epsilon_{\lambda\alpha}^{\;\;\;\;\mu}h_{\mu\nu}$$.}
and the last line is true for the metric (\ref{selfmetric}). So,
for the self-dual warped $AdS_3$ metric a symmetric rank s field
automatically satisfies the tracelessness condition i.e we have
$g^{\mu\nu}\Phi_{\nu\mu\nu_3\cdots\nu_s}=0$. Therefore, the
second line in (\ref{tracelessness}) is equal to zero provided
that the gauge constraint be also equal to zero.

But, for gauge constraint, one has
 \bea
  -m \nabla^\mu \Phi_{\mu \nu_2
\nu_3 \cdots \nu_s} &=&
\epsilon^{\mu\alpha\beta}\nabla_\mu\nabla_\alpha \Phi_{\beta\nu_2
\nu_3 \cdots \nu_s} =  \frac{1}{2}
\epsilon^{\mu\alpha\beta}[\nabla_\mu, \nabla_\alpha]
\Phi_{\beta\nu_2\nu_3\cdots \nu_s} \\ \nonumber &=&
\epsilon^{\mu\alpha\beta} g^{\rho\rho'} R^0_{\beta\rho\mu\alpha}
\Phi_{\rho'\nu_2\nu_3 \cdots \nu_s} +
\epsilon^{\mu\alpha\beta}g^{\rho\rho'}  R^0_{\nu_2
\rho\mu\alpha}\Phi_{\beta\rho'\cdots \nu_s} + \cdots
\\ \nonumber &+&
\epsilon^{\mu\alpha\beta} g^{\rho\rho'} r_{\beta\rho\mu\alpha}
\Phi_{\rho'\nu_2\nu_3 \cdots \nu_s} +
\epsilon^{\mu\alpha\beta}g^{\rho\rho'}  r_{\nu_2
\rho\mu\alpha}\Phi_{\beta\rho'\cdots \nu_s} + \cdots.
 \nonumber
 \eea
As it is obvious, the usual harmonic gauge constraint does not
satisfied for all components of a general rank s field $\Phi_{\mu
\nu_2 \nu_3 \cdots \nu_s}$. In particular, for a spin 2 field
$h_{\mu\nu}$ one obtains
 \bea\label{gc2}
-m \nabla^\mu h_{\mu\theta} &=&0\\
-m \nabla^\mu h_{\mu t} &=&-2(1-\lambda^2)g^{xx}h_{x\theta}\\
-m \nabla^\mu h_{\mu x}
&=&2(1-\lambda^2)(g^{tt}h_{t\theta}+g^{t\theta}h_{\theta\theta}).
 \eea
Thus, the symmetric condition on $h_{\mu\nu}$ at least implies
that $h_{x\theta}=0$. But, from (\ref{leom}) we have
 \bea\label{xtheta}
h_{x\theta}=\frac{g_{xx}}{-m\sqrt
{-g}+g_{xx}\Gamma^x_{t\theta}}(\partial_t
h_{\theta\theta}-\partial_{\theta}h_{t\theta})
 \eea
If we consider the ansatz
\begin{equation}\label{ansatz}
h_{\mu\nu}(x,\theta,\phi)=e^{-i(\omega t-k\theta) }R_{\mu\nu}(x)
\end{equation}
then $h_{x\theta}=0$ means that $\omega=k=0$ which is not
interesting generally. So, we consider in the rest of the paper
the field $h_{\mu\nu}$ is not totally symmetric but is traceless
in the sense that
 \bea\label{newtraceless}
g^{rr}h_{rr}+g^{\theta\theta}h_{\theta\theta}+g^{t\theta}(h_{t\theta}+h_{\theta
t})+g^{tt}h_{tt}=0
 \eea
This implies that we should impose following conditions
 \bea\label{newconditions}
\nabla_{i}(h_{j r}-h_{rj})&=&0,
 \eea
 where $\{i,j\}=\{t,\theta\}$.
To proceed we also consider one of the gauge constraint
 \bea\label{gaugetheta}
\nabla^\mu h_{\mu\theta} =0,
 \eea
yet can be satisfied. This help us to simplify and write the
equation of motion for $h_{\theta\theta}$ and $h_{t\theta}$ in a
simple form. In fact, the three gauge constraint and one
tracelessness condition in symmetric case replaced to three
conditions (\ref{newconditions}) and a gauge constraint
(\ref{gaugetheta}). The above conditions can be generalized for a
higher spin as following
 \bea\label{constraint}
\nabla_{i}(\Phi_{j\cdots r\cdots}-\Phi_{r\cdots j\cdots})&=&0,\\
g^{\alpha\beta}\Phi_{\nu_1\cdots\alpha\cdots\beta\cdots\nu_s}&=&0,\\
\nabla^{\mu}\Phi_{\mu\theta\cdots\theta}&=&0.
 \eea
Let us look for a Klein-Gordon equation for fields. From
(\ref{leom}) one finds
\begin{eqnarray}
\label{neom}
 (\nabla ^2 -m^2)  \Phi _{\mu\nu_2 \nu_3 \cdots \nu_s} &=& \nabla ^\sigma \nabla _\mu  \Phi _{\sigma\nu_2 \nu_3 \cdots
 \nu_s}\nonumber \\
&=& g^{\sigma\rho} [\nabla_\rho,\nabla_\mu]\Phi_{\sigma\nu_2
\nu_3 \cdots \nu_s}+ \nabla_{\mu} \nabla ^\sigma  \Phi
_{\sigma\nu_2 \nu_3 \cdots
 \nu_s} \nonumber \\
&=&g^{\sigma\rho} g^{\eta\delta}
(R_{\sigma\delta\rho\mu}\Phi_{\eta\nu_2 \nu_3 \cdots
\nu_s}+R_{\nu_2\delta\rho\mu}\Phi_{\sigma\eta\nu_3\cdots\nu_s}
\nonumber \\
 &&
\;\;+R_{\nu_3\delta\rho\mu}\Phi_{\sigma\nu_2\eta\nu_4\cdots\nu_s}+\cdots
+R_{\nu_s\delta\rho\mu}\Phi_{\sigma\nu_2\nu_3\cdots\nu_{s-1}\eta}) \nonumber \\
&&\;\; +\nabla_{\mu} \nabla ^\sigma  \Phi _{\sigma\nu_2 \nu_3
\cdots \nu_s}\nonumber\\
 &=& -\frac{\lambda^2}{4}(s+1)\Phi_{\mu\nu_2 \nu_3 \cdots \nu_s}
\nonumber \\
&+& g^{\sigma\rho} g^{\eta\delta}
(r_{\sigma\delta\rho\mu}\Phi_{\eta\nu_2 \nu_3 \cdots
\nu_s}+r_{\nu_2\delta\rho\mu}\Phi_{\sigma\eta\nu_3\cdots\nu_s}
\nonumber \\
&&
\;\;+r_{\nu_3\delta\rho\mu}\Phi_{\sigma\nu_2\eta\nu_4\cdots\nu_s}+\cdots
+r_{\nu_s\delta\rho\mu}\Phi_{\sigma\nu_2\nu_3\cdots\nu_{s-1}\eta})\nonumber\\
&+&\nabla_{\mu} \nabla ^\sigma  \Phi _{\sigma\nu_2 \nu_3 \cdots
\nu_s}.
\end{eqnarray}
Especially, for $\Phi _{\theta \theta \cdots \theta}$ and $\Phi
_{t \theta \cdots \theta}$ we obtain
 \bea\label{eomttheta}
(\nabla ^2 -m^2)  \Phi _{\theta \theta \cdots
\theta}&=&-\frac{\lambda^2}{4}(s+1)\Phi_{\theta\theta\cdots\theta},\\
(\nabla ^2 -m^2)  \Phi _{t \theta \cdots
\theta}&=&-\frac{\lambda^2}{4}(s+1)\Phi_{t\theta\cdots\theta}+(1-\lambda^2)g^{xx}g^{t\eta}\Phi_{\eta\theta\cdots\theta}.
 \eea
\section{Rank 2 Field in Self-dual Warped AdS3}
In this section, we consider a rank 2 tensor field and study its
dynamics in self dual warped $AdS3$ background. We solve the
equations of motion and then find the quasi normal modes. At
last, we will do the computation for the extremal case.

\subsection{Field Dynamics}
Here, we consider a traceless rank 2 field $h_{\mu\nu}$ which
should satisfy (\ref{leom}). Here, for later use, we present this
equation in details using the notation of \cite{Chen:2010ik} as
following
 \bea\label{leomdetails}
\partial_xh_{\theta\theta}&=&\partial_{\theta}h_{r\theta}+\Gamma_{(\theta\theta)}+m_{(\theta\theta)},\cr
\partial_xh_{t\theta}&=&\partial_{t}h_{r\theta}+\Gamma_{(t\theta)}+m_{(t\theta)},\cr
\partial_xh_{\theta
t}&=&\partial_{\theta}h_{rt}+\Gamma_{(\theta t )}+m_{(\theta
t)},\cr
\partial_xh_{tt}&=&\partial_{t}h_{rt}+\Gamma_{(tt)}+m_{(tt)},\cr
-mh_{x\theta}&=&g_{xx}\epsilon^{xt\theta}(\partial_{t}h_{\theta\theta}-\partial_{\theta}h_{t\theta})+\Gamma_{(x\theta)},\cr
-mh_{xt}&=&g_{xx}\epsilon^{xt\theta}(\partial_{t}h_{\theta t
}-\partial_{\theta}h_{tt})+\Gamma_{(xt)},\cr
-mh_{xx}&=&g_{xx}\epsilon^{xt\theta}(\partial_{t}h_{\theta x
}-\partial_{\theta}h_{tx})+\Gamma_{(xx)},
 \eea
where
 \bea
\Gamma_{(\theta\theta)}&=&\Gamma^{\lambda}_{r\theta}h_{\theta\lambda}-\Gamma^{\lambda}_{\theta\theta}h_{r\lambda},\cr
\Gamma_{(t\theta)}&=&\Gamma^{\lambda}_{r\theta}h_{t\lambda}-\Gamma^{\lambda}_{t\theta}h_{r\lambda},\cr
\Gamma_{(\theta t)
}&=&\Gamma^{\lambda}_{rt}h_{\theta\lambda}-\Gamma^{\lambda}_{\theta
t }h_{r\lambda},\cr
\Gamma_{(tt)}&=&\Gamma^{\lambda}_{rt}h_{\theta\lambda}-\Gamma^{\lambda}_{t
t }h_{r\lambda},\cr
 \Gamma_{(x\theta)}&=&g_{xx}\epsilon^{xt\theta}(-\Gamma^{\lambda}_{t\theta}h_{\theta\lambda}+\Gamma^{\lambda}_{\theta\theta}h_{t\lambda}),\cr
 \Gamma_{(xt)}&=&g_{xx}\epsilon^{xt\theta}(-\Gamma^{\lambda}_{tt}h_{\theta\lambda}+\Gamma^{\lambda}_{\theta
 t}h_{t\lambda}),\cr
 \Gamma_{(xx)}&=&g_{xx}\epsilon^{xt\theta}(-\Gamma^{\lambda}_{tx}h_{\theta\lambda}+\Gamma^{\lambda}_{\theta x}h_{t\lambda}),
\\ \cr
m_{(\theta\theta)}&=&-mg_{xx}\epsilon^{\theta
rt}(g_{t\theta}h_{\theta\theta}-g_{\theta\theta}h_{t\theta}),\cr
m_{(t\theta)}&=&-mg_{xx}\epsilon^{tr\theta}(g_{t\theta}h_{\theta
t }-g_{tt}h_{\theta\theta}),\cr m_{(\theta
t)}&=&-mg_{xx}\epsilon^{\theta rt}(g_{t\theta}h_{\theta t
}-g_{\theta\theta}h_{tt}),\cr
m_{(tt)}&=&-mg_{xx}\epsilon^{tr\theta}(g_{t\theta}h_{tt}-g_{tt}h_{\theta
t}).
 \eea

Next, we focus on the equations of $h_{\theta\theta}$ and
$h_{t\theta}$ components which decouple from the other components
and can be written as (\ref{eomttheta}). Although, for finding
the quasi-normal modes we should also find the exact solution for
$h_{tt}$ but,  using (\ref{leom}), we will able to find exact
solutions for $h_{\theta t}, h_{tt}, h_{x\theta}, h_{xt}$ and
$h_{xx}$.

Recalling the fact that $\Gamma^{\lambda}_{\theta\theta}=0$ and
the ansatz (\ref{ansatz}) and using the constraint
(\ref{constraint}), one can find the exact solution as following.
Firstly, one should solve (\ref{eomttheta}) to find
$h_{\theta\theta}$ and $h_{t\theta}$. Then, from the first two
equations of (\ref{leomdetails}) one can find $h_{\theta t}$ and
$h_{r\theta}$. After that, from the last three equations of
(\ref{leomdetails}) one finds $h_{tr}, h_{r\theta}$ and especially
$h_{rt}$ in terms of $h_{\theta\theta}, h_{t\theta}, h_{\theta
t}$ and $h_{tt}$. Finally, inserting the $h_{rt}$ in the third
equation of (\ref{leomdetails}) one can obtain $h_{tt}$. In the
following we present the results of the computations.
\subsection{Solution for $h_{\theta\theta}$ and $h_{t \theta}$}
The equations of motion for $h_{\theta\theta}$ and $h_{t\theta}$
reads as
 \bea
(\nabla ^2 -m^2)
\Phi_{\theta\theta}&=&-3\frac{\lambda^2}{4}\Phi_{\theta\theta},\\
(\nabla ^2 -m^2)  \Phi
_{t\theta}&=&(\frac{\lambda^2}{4}-1)\Phi_{t\theta}+(1-\lambda^2)(x-\frac{x_++x_-}{2})h_{\theta\theta}.
 \eea
So, first of all, we should evaluate $\nabla^2$. As
\cite{Datta:2011za}, one has
\begin{eqnarray}\label{laplacian}
\nabla^2 h_{\mu\nu}&=&\Delta h_{\mu\nu}\nonumber\\&-&
\frac{1}{\sqrt{-g}}\partial_{\alpha}(\sqrt{-g}g^{\alpha\beta}
\Gamma ^\sigma_{\beta\mu} ) h_{\sigma\nu}
 -\frac{1}{\sqrt{-g}}\partial_{\alpha}(\sqrt{-g}g^{\alpha\beta}
\Gamma ^\sigma_{\beta\nu} ) h_{\mu\sigma} \nonumber
\\
&-&2\Gamma^\rho_{\alpha\mu}g^{\alpha\beta}\nabla_\beta h_{\rho\nu}
-2\Gamma^\rho_{\alpha\nu}g^{\alpha\beta}\nabla_\beta h_{\mu\rho}\nonumber\\
&-&g^{\alpha\beta}\Gamma^\sigma_{\beta\mu}\Gamma^\rho_{\alpha\sigma}
h_{\rho\nu}
-g^{\alpha\beta}\Gamma^\sigma_{\beta\nu}\Gamma^\rho_{\alpha\sigma}
h_{\mu\rho}\nonumber\\
&-&g^{\alpha\beta}\Gamma^\rho_{\alpha\mu}\Gamma^\sigma_{\beta\nu}
h_{\sigma\rho}
-g^{\alpha\beta}\Gamma^\rho_{\alpha\mu}\Gamma^\sigma_{\beta\nu}
h_{\rho\sigma}
\end{eqnarray}
where $\triangle$ is the usual scalar Laplacian
\begin{equation}\label{slaplacian}
\Delta h_{\mu\nu}=\frac{1}{\sqrt{-g}}\partial_\alpha
(\sqrt{-g}g^{\alpha \beta} \partial_\beta h_{\mu\nu}).
\end{equation}
Calculations of the first and second line of (\ref{laplacian}) are
straightforward. For the third line, one can see in a background
where its metric are only x-dependent and $g_{xi}=0$, only
$\nabla_{i}h_{xj}$ and $h_{xx}$ contribute to
$\nabla^2h_{ij}$($\{i,j\}=t,\theta$). Using the first order
equation (\ref{leom}) and the constraint (\ref{newconditions}),
one can rewrite the third line in terms of $h_{\theta\theta},
h_{t\theta}, h_{tt}$. For example
 \begin{eqnarray}\label{WB5}
 \nabla_t h_{xt}&=&m\frac{\sqrt{-g}}{-\hat{g}}(-g_{t\theta}h_{tt}+g_{tt}h_{t\theta})+\nabla_x
h_{tt},
\end{eqnarray}
where $\hat{g}=g_{tt}g_{\theta\theta}-(g_{t\theta})^2$.

Moreover, by using (\ref{WB5}) one can show that the first and
second terms in the forth line of (\ref{laplacian}) are equal to
zero and just the last expression would contribute to $\nabla^2$.
In this expression we have $h_{xx}$ term which can be replaced
using traceless condition. Using the above equations and inserting
the ansatz (\ref{ansatz}) one can obtain the following set of
differential equations for $h_{\theta\theta}$ and $h_{t\theta}$
 \bea\label{equation2}
 (\Delta \textbf{1}_{2\times 2}+M_{2\times 2})
\begin{pmatrix}
R_{t\theta}\\
R_{\theta\theta}
\end{pmatrix}=0,
 \eea
where the mass matrix $M$ is given by
 \bea
M_{2\times 2}=\begin{pmatrix}
-m^2+\frac{\lambda^2}{4}&(1-\frac{2m}{\lambda})(1-\lambda^2)(x-x_0) \\
0 &-(m-\lambda)^2+\frac{\lambda^2}{4}
\end{pmatrix}
 \eea
and the operator $\Delta$ is equal to
 \bea
 \Delta =(x-x_+)(x-x_-)\frac{\partial^2}{\partial
x^2}+2(x-x_0)\frac{\partial}{\partial x}+Q(x),
 \eea
where
 \bea\label{qux}
Q(x)=\frac{(\omega+k(x-x_0))^2}{(x-x_+)(x-x_-)}-\frac{k^2}{\lambda^2}
 \eea
As it is clear, except for the $h_{\theta\theta}$, the equations
(\ref{equation2}) are a set of nonlinear coupled equations. For
solving such coupled equations, one may try to diagonalize  $M$
and decouple the equations but, the components of M are
x-dependent and we should be careful for doing such procedure.
Overcoming such problem is easy by defining the following new
fields
 \bea
H_{t\theta}&=&\frac{R_{t\theta}}{(x-x_0)}\\
H_{\theta\theta}&=&R_{\theta\theta}.
 \eea
Then, one can obtain the new set of  equations as
\bea\label{equation2}
 (\delta_{2\times 2}+{\cal M}_{2\times 2})
\begin{pmatrix}
H_{t\theta}\\
H_{\theta\theta}
\end{pmatrix}=0,
 \eea
where the new matrix ${\cal M}$ is a constant matrix and is given
by
$$ {\cal M}_{ij}=\frac{M_{ij}}{(x-x_0)^{j-i}}.$$
Also, the new diagonal differential operator $\delta$ is as
 \bea
\delta_{11}&=&\Delta
 +2\frac{(x-x_+)(x-x_-)}{(x-x_0)}\frac{\partial}{\partial
 x}+2,\\
 \delta_{22}&=&\Delta,
 \eea
and all other components of $\delta$ are zero. Now, we can
diagonalize ${\cal M}$ and find new decoupled equations. Noting
that the eigenvalues of $\cal M$ are ${\cal M}_{11}$ and $ {\cal
M}_{22}$, one can find the eigenvector  and matrix transformation
$U$ such that $U{\cal M}U^{-1}$ becomes diagonal. The new fields
in which we have decoupled differential equations are ${\cal
H}=UH$ where
 \bea
 {\cal H}_{t\theta}&=&H_{t\theta}-\frac{1-\lambda^2}{\lambda^2}H_{\theta\theta}\\
 {\cal H}_{\theta\theta}&=&H_{\theta\theta}.
 \eea
Doing the above prescription and using the following common change
of variable
\begin{equation}\label{change}
z=\frac{x-x_+}{x-x_-}
\end{equation}
one finally obtains
 \bea
z(1-z)\frac{\partial^2}{\partial z^2}{\cal
H}_{t\theta}&+&\left(1-z+\frac{4z}{1+z}\right)\frac{\partial}{\partial
z}{\cal H}_{t\theta}\hspace{5.5cm}\nonumber\\
&+&\left(Q(z)+\frac{2}{1-z}+\frac{{\cal M}_{22}}{1-z}\right){\cal
H}_{t\theta}=0,\\
z(1-z)\frac{\partial^2}{\partial z^2}{\cal H}_{\theta\theta}&+&
\left(1-z \right)\frac{\partial}{\partial
z}{\cal H}_{\theta\theta}\hspace{5.5cm}\nonumber\\
&+&\left(Q(z)+\frac{{\cal M}_{33}}{1-z}\right){\cal
H}_{\theta\theta}=0,
 \eea
 where $$Q(z)=-\frac{(\omega-\frac{x_+-x_-}{2}k)^2}{(x_+-x_-)^2}
+\frac{(\omega+\frac{x_+-x_-}{2}k)^2}{(x_+-x_-)^2}\frac{1}{z}+\frac{k^2(1-\frac{1}{\lambda^2})}{1-z}.$$
The solutions of the above equations can be written in terms of
hypergeometric function as following
 \bea\label{solution}
 {\cal
H}_{\theta\theta}(z)&=&
z^{\alpha}(1-z)^{\beta_{\theta\theta}+1}\left(C_{\theta\theta}
F(a_{\theta\theta}+1,b_{\theta\theta+1},c;z))\right .\cr
&&\hspace{3cm}\left . +D_{\theta\theta}
F(a_{\theta\theta}^*+1,b_{\theta\theta}^*+1,c^*;z\right),\\
{\cal
H}_{t\theta}(z)&=&z^{\alpha}(\frac{1-z}{1+z})(1-z)^{\beta_{t\theta}+1}\left(
C_{t\theta}F(a_{t\theta}+1,b_{t\theta}+1,c;z)\right .\cr
&&\hspace{4cm}\left .+D_{t\theta}
F(a_{t\theta}^*+1,b_{t\theta}^*+1,c^*;z)\right),
 \eea
 where $C_{ij}, D_{ij}$
 are arbitrary constants and other parameters are given by
 \newpage
 \bea\label{parameters}
&&\alpha=-\frac{i}{2}(k+\frac{2\omega}{x_+-x_-})\;\;\;\;\;\;\;\;\;\;c=1+2\alpha,\cr
&&a_{ij}=\beta_{ij}-ik,\;\;\;\;\;\;b_{ij}=\beta_{ij}-\frac{2i\omega}{x_+-x_-},\cr
&&\beta_{t\theta}^{\pm}=-\frac{1}{2}\pm\frac{1}{2}\sqrt{1-4k^2(1-\frac{1}{\lambda^2})-4M_{11}},\cr
&&\beta_{\theta\theta}^{\pm}=-\frac{1}{2}\pm\frac{1}{2}\sqrt{1-4k^2(1-\frac{1}{\lambda^2})-4
M_{22}}.
 \eea
We should mention that in the next section when we will impose the
Dirichlet boundary condition, for finding a suitable solution, we
will choose the plus sign for $h_{t \theta}$ and
$h_{\theta\theta}$.
\subsection{Solution for $h_{\theta t}$ and $h_{tt}$}
As we mentioned before, for finding $h_{\theta t}$ and $h_{tt}$,
firstly, we solve the (\ref{eomttheta}) to find $h_{\theta\theta}$
and $h_{t\theta}$. Then, from the first two equations of
(\ref{leomdetails}) we obtain $h_{\theta t}$ and $h_{r\theta}$.
After that, from the last three equations of (\ref{leomdetails})
we obtain $h_{tr}, h_{r\theta}$ and especially $h_{rt}$ in terms
of $h_{\theta\theta}, h_{t\theta}, h_{\theta t}$ and $h_{tt}$.
Finally, by inserting the $h_{rt}$ in the third equation of
(\ref{leomdetails}) we obtain $h_{tt}$. Notice that we use the
ansatz (\ref{ansatz}). The solution may be written as
 \bea\label{solution2}
 h_{\theta t}&=&\frac{k}{A}\frac{d}{dx}h_{t\theta}+\frac{\omega}{A}\frac{d}{dx}h_{\theta\theta}-\frac{\omega B+kC}{A}h_{t \theta}
 -\frac{\omega D+kE}{A}h_{\theta\theta}
 ,\cr
 h_{tt}&=&\frac{1}{\bar{A}}\frac{d}{dx}h_{\theta t}+\frac{F}{\bar{A}}\frac{d}{dx}h_{\theta\theta}-\frac{G}{\bar{A}}h_{\theta t}
  -\frac{H}{\bar{A}}h_{t\theta }-\frac{I}{\bar{A}}h_{\theta\theta},
 \eea
where
 \bea
A&=&\omega\Gamma^{t}_{x\theta}+k\left(\Gamma^{x}_{t\theta}g_{xx}g^{t\theta}-m\epsilon^{tr\theta}g_{xx}g_{t\theta}\right),\cr
\bar{A}&=&\frac{g_{xx}}{m}\epsilon^{xt\theta}\left(-k^2+\Gamma^{x}_{t\theta}\Gamma^{t}_{x\theta}+m^2g_{\theta\theta}\right),\cr
B&=&m\epsilon^{\theta xt}g_{xx}g_{\theta\theta},\cr
 C&=&\Gamma^{\theta}_{x\theta}+g_{xx}g^{t\theta}\Gamma^{x}_{t\theta},\cr
 D&=&\Gamma^{\theta}_{x\theta}-m\epsilon^{\theta
 rt}g_{xx}g_{t\theta},\cr
 E&=&g_{xx}g^{\theta\theta}\Gamma^x_{t\theta}+m\epsilon^{tx\theta}g_{xx}g_{tt},\cr
 F&=&-\frac{\omega}{k}-\frac{\Gamma^x_{tt}}{\Gamma^x_{t\theta}},\cr
 G&=&\Gamma^t_{xt}-F\Gamma^{t}_{x\theta}-\frac{g_{xx}}{m}\epsilon^{xt\theta}\left(\omega k+\Gamma^{x}_{t\theta}\Gamma^t_{xt}
 +m^2g_{t\theta}\right),\cr
 H&=&\frac{g_{xx}}{m}\epsilon^{xt\theta}\left(F(k^2-m^2\sqrt{-g}g_{\theta\theta})+\Gamma^{x}_{t\theta}\Gamma^{\theta}_{x\theta}\right),\cr
 I&=&\frac{g_{xx}}{m}\epsilon^{xt\theta}\left(F(\omega k-\gamma^{\theta}_{x\theta}+
 m^2\sqrt{-g}g_{\theta t})+\Gamma^{x}_{t\theta}\Gamma^{\theta}_{xt}\right)+\Gamma^{\theta}_{xt}.
 \eea
 Using the solution (\ref{solution}) and the following contiguous relations between hypergeometric functions
 \bea\label{rec}
\frac{\partial}{\partial
z}F_1(a,b,c;z)=\frac{ab}{c}F_1(a+1,b+1,c+1;z),\hspace{5.1cm}\nonumber\\
azF(a+1,b+1,c+1;z)=cF(a,b+1,c;z)-cF(a,b,c;z),\hspace{2.3cm}\nonumber\\
a(1-z)F(a+1,b,c;z)=(c-b)F(a,b-1,c;z)-(c-a-b)F(a,b,c;z),
 \eea
 one can rewrite the above solutions in terms of ordinary hypergeometric
 functions. This calculation is straightforward but tedious. We
 present the result of such calculation for the case where the
 fields are completely symmetric in Appendix A.
\subsection{Quasi Normal Modes}
Quasi normal modes can be found by imposing Dirichlet boundary
condition on the solutions (\ref{solution}) and (\ref{solution2}).

Before that, we mention some points. Firstly, because our aim is
to find the quasi-normal modes, we consider in-going waves into
horizon and so we choose the constant $D_{\theta\theta}=0$ and
$D_{t\theta}=0$. Secondly, we also choose the plus sign
(\ref{solution}). The calculations for minus sign is similar to
the case of plus sign. Thirdly, one can see that when $\lambda=1$
then the solution (\ref{solution}) becomes to the solution that
was found in \cite{Datta:2011za}. In fact, this is due to the fact
that at $\lambda=1$ the geometry of self-dual warped $AdS_3$
becomes the usual $BTZ$ geometry. So, we can use the result of
computations was done in \cite{Datta:2011za} for finding ratio of
the coefficients $C_{\theta\theta}$ and $C_{t\theta}$. So, all
coefficients of solutions of all fields can be written in terms of
$C_{\theta\theta}$. Now, because of the reason which will soon be
present, let us choose the $C_{\theta\theta}$ as following
 \bea
C_{\theta\theta}=C_0(a_{\theta\theta})(a_{\theta\theta}-1),
 \eea
where $C_0$ is an arbitrary constant independent of
$a_{\theta\theta}$ and $b_{\theta\theta}$.

After all, using the following transformation relation between
hypergeometric functions
 \bea\label{asymptotic}
&&F(a,b,c;z)=\frac{\Gamma(c)\Gamma(c-a-b)}{\Gamma(c-a)\Gamma(c-b)}F(a,b,a+b-c+1;1-z)+\cr
&&+(1-z)^{c-a-b}\frac{\Gamma(c)\Gamma(a+b-c)}{\Gamma(a)\Gamma(b)}F(c-a,c-b,c-a-b+1;1-z)
 \eea
one can find the asymptotic behavior of the solution where
$z\rightarrow 1$. Choosing the plus sign in (\ref{solution})
which means that the leading order of solutions comes from the
second line of (\ref{asymptotic}), one can easily find the
asymptotic behavior of all terms of solution of all fields. Let
us focus on one of these terms in $h_{tt}$ solution. That is
 \bea
h_{tt}=\frac{1}{\bar{A}}\frac{d}{dx}h_{\theta
t}+\cdots=\frac{1}{\bar{A}}\frac{d}{dx}(\frac{\omega}{A}\frac{d}{dx}h_{\theta
\theta})+\cdots.
 \eea
Using the change of variable (\ref{change}), it is not hard see
that this term may be written as
 \bea
&&\frac{1}{\bar{A}}\frac{d}{dx}(\frac{\omega}{A}\frac{d}{dx}h_{\theta
\theta})=C(z)z^2\frac{d^2}{dz^2}h_{\theta\theta}+\cdots\cr
 &&\sim C(z)(a_{\theta\theta})(a_{\theta\theta}-1)z^{\alpha+2}(1-z)^{-\beta_{\theta\theta}-2}
 \frac{\Gamma(c)\Gamma(a_{\theta\theta}+b_{\theta\theta}+2-c)}{\Gamma(a_{\theta\theta}+1)\Gamma(b_{\theta\theta}+1)}+\cdots,
 \eea
 where $C(z)$ is a function of $z$ and goes to a finite constant
 when $z\rightarrow 1$.
Now, imposing Dirichlet condition on $h_{tt}$ implies that
 \bea
a+1-2=-n_1,\;\;\;\;\;\;\;or\;\;\;\;\;\;b+1=-n_2.
 \eea
where $n_1$ and $n_2$ are non-negative integers. One can see that
these conditions are \textit{necessary} and \textit{sufficient}
for satisfying Dirichlet condition on all fields.

Finally, one can obtain the left and right quasi
 normal modes as
 \bea
k=-i2\pi T_Ll(n_1+h_L),\;\;\;\;\;\;\;\;\;\omega=-i2\pi
T_Rl(n_2+h_R),
 \eea
 where
 \bea\label{LR}
&&h_R=+\frac{1}{2}+\frac{1}{2}\sqrt{1-4k^2(1-\frac{1}{\lambda^2})-4
M_{22}}\cr
&&h_L=-\frac{3}{2}+\frac{1}{2}\sqrt{1-4k^2(1-\frac{1}{\lambda^2})-4
M_{22}}.
 \eea
Notice that from (\ref{LR}) one has $h_R-h_L=+2$. Note also that
if one chooses
$C_{\theta\theta}=C_0(b_{\theta\theta})(b_{\theta\theta}-1)$ then
will find $h_R-h_L=-2$.

The above result for conformal weight are in precise agreement
with the calculations in [hidden] where the conformal weight
obtained using an algebraic way from highest-weight mode.

At the end, one can also do the above calculations for the
extremal warped $AdS_3$ black hole. The solution of equation of
motion are presented in appendix B.
\section{Higher Spin on Self-dual Warped $AdS_3$}
In this section, we discuss how we can find the conformal weight
and  quasi normal modes of a field with arbitrary spin in self
dual warped $AdS_3$ background. The Key point is that, as for the
spin 2 case, for finding the conformal weight and quasi-normal
modes of higher spin fields, it is sufficient to find the
solution of $h_{\theta\theta...\theta}$ equation of motion. So,
let us firstly find this solution in the next section.

\subsection{Solution for $\Phi_{\theta\theta\cdots\theta}$}
The equation of motion for $\Phi_{\theta\theta\cdots\theta}$ is
given by (\ref{eomttheta}). So, we should evaluate $\nabla^2$ as
follows\footnote{In this section, by expressions similar to $
\Gamma^{\sigma}_{\mu\nu}\Phi_{\cdots\sigma\cdots}$ we mean that
$\Gamma^{\sigma}_{\mu\nu}\phi_{\sigma\nu_2\cdots\nu_s}+\Gamma^{\sigma}_{\mu\nu}\phi_{\nu_1\sigma\cdots\nu_s}
+\cdots+\Gamma^{\sigma}_{\mu\nu}\phi_{\nu_1\nu_2\cdots\sigma}$.}
\begin{eqnarray}\label{LLaplacian}
\nabla^2 \Phi_{\theta\theta\cdots\theta}= \Delta
\Phi_{\theta\theta\cdots\theta}&-&\frac{1}{\sqrt{-g}}(\partial_{\alpha}\sqrt{-g}g^{\alpha\beta}
\Gamma_{\alpha\theta}^{\sigma})\Phi_{\theta\cdots\sigma\cdots\theta}\nonumber\\
&-&2g^{\alpha\beta}\Gamma^\sigma_{\alpha\theta}\nabla_\beta
\Phi_{\theta\cdots\sigma\cdots\theta},\nonumber\\
&-&g^{\alpha\beta}\Gamma^\rho_{\beta\theta}\Gamma^\sigma_{\alpha\rho}
\Phi_{\theta\cdots\sigma\cdots\theta}\nonumber\\
&-&g^{\alpha\beta}\Gamma^\sigma_{\alpha\theta}\Gamma^\rho_{\beta\theta}
\Phi_{\theta\cdots\sigma\cdots\rho\cdots\theta}
\end{eqnarray}
where the $\Delta$ is the scalar Laplacian (\ref{slaplacian}).

Again, using the first order equation (\ref{leom}) and the
constraint (\ref{constraint}), we are able to rewrite the
$\nabla_{i}\Phi_{\nu_1\cdots j\cdots\nu_s}$ and $\Phi_{\nu_1\cdots
x\cdots x\cdots \nu_s}$ in terms of
$\Phi_{\theta\theta\nu_3\cdots\nu_s},
\Phi_{t\theta\nu_3\cdots\nu_s}, \Phi_{\theta t \nu_3\cdots\nu_s}$
and $\Phi_{tt\nu_3\cdots\nu_s}$ without any x indices.

%

So, using the ansatz
$$\Phi_{\theta\theta\cdots\theta}=e^{-i(\omega
t-k\theta)}R_{\theta\theta\cdots\theta}$$ the final differential
equation for $\Phi_{\theta\theta\cdots\theta}$ reads as
 \bea
z(1-z)\frac{\partial^2}{\partial
z^2}R_{\theta\theta\cdots\theta}+\left(1-z\right)\frac{\partial}{\partial
z} R_{\theta\theta\cdots\theta}+\left(Q(z)+\frac{
\widetilde{M}_{22}}{1-z}\right)R_{\theta\theta\cdots\theta}=0,
 \eea
where we have used (\ref{change}) and
 \bea
\widetilde{M}_{22}=-(m-s\frac{\lambda}{2})^2+\frac{\lambda^2}{4}.
 \eea
The solution of the above equation with in-going condition on
horizon is
 \bea
R_{\theta\theta\cdots\theta}=\widetilde{C}_{\theta\theta}z^{\alpha}(1-z)^{\widetilde{\beta}_{\theta\theta}+1}
F_1(\widetilde{\beta}_{\theta\theta}-ik+1,\widetilde{\beta}_{\theta\theta}-\frac{2i\omega}{x_+-x_-}+1,c;z)
 \eea
 where
 \bea
 \widetilde{\beta}_{\theta\theta}=-\frac{1}{2}+\frac{1}{2}\sqrt{1-4k^2(1-\frac{1}{\lambda^2})-4
\widetilde{M}_{22}},
 \eea
 and $\alpha,c$ are given in (\ref{parameters}).
\subsection{Quasi-Normal Modes}
 For imposing Dirichlet boundary conditions on all
fields, as for spin 2 case, we focus on $\Phi_{tt\cdots t}$. The
equation involving $\Phi_{tt\cdots t}$ is
 \bea
\partial_x\Phi_{\theta t\cdots
t}=\partial_{\theta}\Phi_{xt\cdots t}+\Gamma_{(\theta t\cdots
t)}+m_{(\theta\ t\cdots t)},
 \eea
where
 \bea\label{tt...}
 \Gamma_{(\theta t\cdots t)}&=&\Gamma^{\lambda}_{r\theta}\Phi_{\theta\cdots \lambda \cdots}-\Gamma^{\lambda}_{\theta t}\Phi_{r\cdots\lambda\cdots},\cr
 m_{(\theta t\cdots t)}&=&-mg_{xx}\epsilon^{\theta
 rt}(g_{t\theta}\Phi_{\theta t\cdots t}-g_{\theta\theta}\Phi_{t\cdots
 t}).
 \eea
In the above expression, one can find the mostly-t field
$\Phi_{tt\cdots t}$ in terms of fields with less t indices. So,
using (\ref{tt...}), among the many terms, one finds a term with
maximum power of z and derivative of
$\Phi_{\theta\theta\cdots\theta}$ as
 \bea
\Phi_{tt\cdots t}\sim
z^{s}\frac{d^s}{dz^s}\Phi_{\theta\theta\cdots\theta}+\cdots.
 \eea
Now, choosing
$$C_{\theta\theta\cdots\theta}=C_{0}(\tilde{a}_{\theta\theta})\cdots(\tilde{a}_{\theta\theta}-s+1)$$
and using the asymptotic behavior of hypergeometric functions and
imposing Dirichlet condition on $\Phi_{t\cdots t}$, one finds the
necessary and sufficient conditions in which all field being zero
at infinity are
 \bea
\tilde{a}_{\theta\theta}+s-1=-n,\;\;\;\;or\;\;\;\;\tilde{b}_{\theta\theta}+1=-n,
 \eea
where here $n$ is a non-negative integer.

and obtain
 \bea
k=-i2\pi T_Ll(n_1+h_L),\;\;\;\;\;\;\;\;\;\omega=-i2\pi
T_Rl(n_2+h_R)
 \eea
 where
 \bea\label{LR...}
&&h_R=+\frac{1}{2}+\frac{1}{2}\sqrt{1-4k^2(1-\frac{1}{\lambda^2})-4
\widetilde{M}_{22}}\cr
&&h_L=-\frac{2s-1}{2}+\frac{1}{2}\sqrt{1-4k^2(1-\frac{1}{\lambda^2})-4
\widetilde{M}_{22}}.
 \eea
 \\
\section{Conclusion}
In this paper, we have studied the dynamics of a rank s field in a
3 dimensional self-dual warped $AdS_3$ background. We have firstly
considered the situation in which the field is totally symmetric.
We discussed that imposing the harmonic gauge constraint and
tracelessness condition may not satisfy for a symmetric rank s
field simultaneously. So, we have considered a general rank s
field which should satisfy equation of motion (\ref{leom}).
Although, the fields do not satisfy the symmetric and
tracelessness conditions but we supposed that all modes satisfy
weaker constraint (\ref{constraint}). This constraint greatly
help us to proceed in calculation. In fact, We were able to
obtain the equations of motion for $h_{\theta\theta}$ and
$h_{t\theta}$ modes(for a rank 2 field) which are coupled
nonlinear differential equations decoupled from the other modes.
By a suitable redefinition of the fields, one can decouple the
equations and find the exact solutions which are in terms of
hypergeometric functions. Having found the solution, we have
found the quasi-normal modes which are in-going waves at horizon
and satisfy Dirichlet boundary condition at infinity.

At the end, we have extended our computations for a general rank s
field and found the quasi-normal modes.

Due to less symmetry of the warped $AdS_3$ geometry, all aspects
of such geometries are unknown and many open problems are yet
unsolved. For example, the physics of geometry, fermions dynamics
on this background(work in progress), its black holes and
perturbations around the vaccume solution, the conformal field
theory dual to this geometry and its dictionary are some
interesting problems which should be studied.
\section{Acknowledgement}
Mohammad A. Ganjali would like to thanks the Kharazmi university
for supporting the paper via grant.
\section*{Appendix}
\appendix
 \section{Solution for Symmetric Case}
 Using the solution $h_{\theta\theta}=z^{\alpha}(1-z)^{\beta+1}F(a+1,b+1,c;z)$
 and
 (\ref{leom}) for a symmetric field $h_{\mu\nu}$ and (\ref{rec}), one may obtain the solution for
 other components
 \bea
 h_{\theta
t}&=&A_1z\frac{dh_{\theta\theta}}{dz}+\frac{A_2+A_3z}{1-z}h_{\theta\theta}\cr
&=&z^{\alpha}(1-z)^{\beta}\frac{1}{A}\left(\sum_{i,j=0}^{1}C_{\theta t}^{ij}F(a+1-i,b+1-j,c;z)\right),\\
h_{tt}&=&B_1z\frac{dh_{\theta
t}}{dz}+\frac{B_2+B_3z}{1-z}h_{\theta
t}+\frac{B_4+B_5z+B_6z^2}{(1-z)^2}h_{\theta\theta}\cr
&=&z^{\alpha}(1-z)^{\beta-1}\frac{1}{A\overline{A}}\left(\sum_{i,j=0}^{2}C_{tt}^{ij}F(a+1-i,b+1-j,c;z)\right),
 \eea
 where
 \bea
 C_{\theta t}^{00}&=&\left(3\beta+2+W\right),\cr
 C_{\theta t}^{01}&=&\left(a-2\beta\right),\cr
 C_{\theta t}^{10}&=&-\frac{(a-2\beta)}{(2\beta)}C_{\theta t}^{11}=-\frac{(2\beta)}{(a)}\left(\alpha-\beta-a-2-B\right),\cr
 \cr\cr
 C_{tt}^{00}&=&\left(C_{t\theta}^{00}(3\beta+1+\overline{W}-2(1-\lambda^2)\widetilde{A})+2(1-\lambda^2)\bar{x}_0A\right),\cr
 C_{tt}^{01}&=&2(a-2\beta)\left(3\beta+1-\lambda m-(1-\lambda^2)\widetilde{A}\right),\cr
 C_{tt}^{10}&=&\frac{(2\beta)}{(a)}\left(\frac{a}{2\beta}C_{t\theta}^{10}(3\beta+\overline{W}
 -2(1-\lambda^2)\widetilde{A})-2(1-\lambda^2)\bar{x}_0A\right .\cr
 &&\hspace{1cm}\left
.-C_{t\theta}^{00}(-\alpha+\beta+a+1+\overline{B}-(1-\lambda^2)\widetilde{A})\right),\cr
 C_{tt}^{02}&=&(a-2\beta+1)(a-2\beta),\cr
 C_{tt}^{20}&=&-\frac{(2\beta)(1-2\beta)}{(a)(a-1)}\left(\frac{a}{2\beta}C_{\theta t}^{10}(\alpha-\beta-a-\overline{B}+(1-\lambda^2)\widetilde{A})
 +(1-\lambda^2)A\widetilde{B}\right)\cr
 C_{tt}^{11}&=&\frac{(a-2\beta)}{(a)}
 \left(\frac{a}{2\beta}C_{\theta t}^{10}(5\beta-1+\overline{B}-2(1-\lambda^2)\widetilde{A})-2(1-\lambda^2)A\widetilde{B} \right .\cr
 &&\hspace{2cm} \left .+(-5\beta+1)(-\alpha+\beta+a+1+\overline{B}-(1-\lambda^2)\widetilde{A})\right)\cr
 C_{tt}^{12}&=&\frac{(a-2\beta+1)(a-2\beta)}{(a)})\left(2\alpha-2\beta-2a-3-B-\overline{B}+(1-\lambda^2)\widetilde{A}\right)\cr
 C_{tt}^{21}&=&-\frac{(1-2\beta)}{(a+1-2\beta)}C_{tt}^{22}=-\frac{(a-2\beta)}{(2\beta)}C_{tt}^{20}.
 \eea
Here, we have defined
 \bea
W&=&-\lambda
m-\frac{\lambda^2}{2},\;\;\;\;\;\;\;\;\;\;\overline{W}=-\lambda
m+\frac{\lambda^2}{2},\cr
A&=&\frac{W^2+k^2}{2W\bar{x}_0},\;\;\;\;\;\;\;\;\;\;\;\;\;\overline{A}=\frac{\overline
{W}^2+k^2}{2\overline{W}\bar{x}_0},\cr
B&=&\frac{W^2\bar{x}_0+k\omega}{2W\bar{x}_0},\;\;\;\;\;\;\;\;\overline{B}=\frac{\overline
{W}^2\bar{x}_0+k\omega}{2\overline{W}\bar{x}_0},\cr
\widetilde{A}&=&\frac{W\overline{W}-k^2}{2W\overline{W}},\;\;\;\;\;\;\;\;\;\;\widetilde{B}=\frac{W\overline
{W}\bar{x}_0-k\omega}{2W\overline{W}}.
 \eea
 and we have used the relations (\ref{parameters}).
\section{Extremal Case}
In this appendix, we consider the extremal self-dual warped
$AdS_3$ black hole. Defining $x_+=x_-=x_0$, the equations of
motion of ${\cal H}_{t\theta}$ and ${\cal H}_{\theta\theta}$ read
as
 \bea\label{extremequ}
&&(x-x_0)^2\frac{\partial^2}{\partial x^2}{\cal
H}_{t\theta}+4\left(x-x_0\right)\frac{\partial}{\partial x}{\cal
H}_{t\theta}+\left(Q(x)+2+M_{11}\right){\cal
H}_{t\theta}=0,\hspace{1.5cm}\nonumber\\
&&(x-x_0)^2\frac{\partial^2}{\partial x^2}{\cal
H}_{\theta\theta}+2\left(x-x_0\right)\frac{\partial}{\partial
x}{\cal H}_{\theta\theta}+\left(Q(x)+M_{22}\right){\cal
H}_{tt}=0.\hspace{1.5cm}
 \eea
where $Q(x)$ given by (\ref{qux}). The above equations can be
written in the form of Whittaker differential equation.
%
Then, the solutions can be written in terms of Whittaker functions
 as following
 \bea
{\cal H}_{t\theta}(x)&=&\left( \widetilde{C}_{t\theta}
WhittakerW(-ik,\frac{1}{2}+\beta_{ij},\frac{2i\omega}{x-x_0})\right .\nonumber\\
&+&\left .\widetilde{D}_{t\theta}
WhittakerM(-ik,\frac{1}{2}+\beta_{ij},\frac{2i\omega}{x-x_0})\right)(\frac{1}{x-x_0}),\\
{\cal H}_{\theta\theta}(x)&=& \left(\widetilde{C}_{\theta\theta}
WhittakerW(-ik,\frac{1}{2}+\beta_{ij},\frac{2i\omega}{x-x_0})\right .\nonumber\\
&+&\widetilde{D}_{\theta\theta} \left .
WhittakerM(-ik,\frac{1}{2}+\beta_{ij},\frac{2i\omega}{x-x_0})\right),
 \eea
 where
 \bea
 Whittaker M(\mu,
\nu;z) &=&
e^{-\frac{1}{2}z}z^{\frac{1}{2}+\nu}hypergeom(\frac{1}{2}+\nu-\mu,
1+2\nu; z), \cr
 WhittakerW(\mu, \nu; z) &=&
e^{-\frac{1}{2}z}z^{\frac{1}{2}+nu}KummerU(\frac{1}{2}+\nu-\mu,
1+2\nu;z).\nonumber
 \eea
 Having found the ${\cal H}_{t\theta}$ and ${\cal
 H}_{\theta\theta}$ and using the solutions (\ref{solution2}) for
 the case $x_+=x_-=x_0$, one can find the solution for all other
 modes.
\section{Generalized Contiguous Relations }
 Using (\ref{rec}), we obtains the generalized version of recursion
relations (\ref{rec}) as
 \bea\label{generalrec}
&&\frac{\partial^n}{\partial
z^n}F(a,b,c;z)=\frac{(a)_n(b)_n}{(c)_n}F(a+n,b+n,c+n;z),\nonumber\\
&&\frac{(a)_n}{(c)_n}z^nF(a+n,b+n,c+n;z)=\nonumber\\
&&\hspace{4cm}=\sum_{m=0}^n
\begin{pmatrix}
n\nonumber\\
m
\end{pmatrix}
(-1)^mF(a,b+n-m,c;z)\nonumber\\
&&(1-z)^nF(a+1,b+1,c;z)=\nonumber\\
&&\sum_{m=0}^{n}\prod_{i=0}^{n-m}\prod_{j=0}^{m}(-1)^m
\begin{pmatrix}
n\\
m
\end{pmatrix}
\frac{\Gamma(a-n)}{\Gamma(a)}(c-b+i-1)(c-b-a-j+1)\nonumber\\&&\hspace{2cm}\times
F_1(a-n+1,b-n+m+1,c;z).
 \eea
 where $(a)_n=(a)(a+1)\cdots(a+n-1)$ and $(a)_0=1$.

\end{document}